# ENHANCED SECURE ALGORITHM FOR MESSAGE COMMUNICATION

Shaik Rasool[1], G. Sridhar[2], K. Hemanth Kumar[3], P. Ravi Kumar[4]

[1]Dept. of Computer Science & Engg, S.C.E.T., Hyderabad, India
shaikrasool@live.in
[2]Dept. of Computer Science & Engg, S.C.E.T., Hyderabad, India
gsridhar@live.in
[3]Dept. of Computer Science & Engg, S.C.E.T., Hyderabad, India
khemanth@live.in
[4]Dept. of Computer Science & Engg, S.C.E.T., Hyderabad, India
pravi@hotmail.co.in

*ABSTRACT*

*This paper puts forward a safe mechanism of data transmission to tackle the security problem of information which is transmitted in Internet. The encryption standards such as DES (Data Encryption Standard), AES (Advanced Encryption Standard) and EES (Escrowed Encryption Standard) are widely used to solve the problem of communication over an insecure channel. With advanced technologies in computer hardware and software, these standards seem not to be as secure and fast as one would like. In this paper we propose a encryption technique which provides security to both the message and the secret key achieving confidentiality and authentication. The Symmetric algorithm used has two advantages over traditional schemes. First, the encryption and decryption procedures are much simpler, and consequently, much faster. Second, the security level is higher due to the inherent poly-alphabetic nature of the substitution mapping method used here, together with the translation and transposition operations performed in the algorithm. Asymmetric algorithm RSA is worldwide known for its high security. In this paper a detailed report of the process is presented and analysis is done comparing our proposed technique with familiar techniques*

## 1. INTRODUCTION

In open networked systems, information is being received and misused by adversaries by means of facilitating attacks at various levels in the communication. The encryption standards such as DES (Data Encryption Standard) [6], AES (Advanced Encryption Standard) [7], and EES (Escrowed Encryption Standard) [8] are used in Government and public domains. With today's advanced technologies these standards seem not to be as secure and fast as one would like. High throughput encryption and decryption are becoming increasingly important in the area of high-speed networking [9].With the ever-increasing growth of multimedia applications, security is an important issue in communication and storage of images, and encryption is one the ways to ensure security. Image encryption has applications in inter-net communication, multimedia systems, medical imaging, telemedicine, and military communication. There already exist several image encryption methods. They include SCAN-based methods, chaos-based methods, tree structure-based methods, and other miscellaneous methods. However, each of them has its strength and weakness in terms of security level, speed, and resulting stream size metrics. We hence proposed the new encryption method to overcome these problems [1].

Communication is a major impact in today's business. The communication devices transmit large amount of data with high security. In business, the amount approximately worth over $1 trillion is being transacted every week on the Net. But, unfortunately, the cyber-crimes are nearly 97% and such crimes are undetected [10]. The security is still remains a risky one. At present, various types of cryptographic the algorithms provide high security in information, computer and network-related activities. These algorithms are required to protect the data, integrity and authenticity from various attacks.

This paper discusses a new technique of encryption algorithm which combines a symmetric algorithm FSET (Fast and Secure Encryption Technique) proposed by Varghese Paul [2] with extra features and asymmetric algorithm RSA with Hash Coding (SHA -2). The FSET algorithm is a direct mapping poly alphabetic Symmetric-key encryption algorithm. Here, direct substitution mapping and subsequent translation and transposition operations using X-OR logic and circular shifts that results in higher conversion speed are used. The block size is 128 bits (16 characters) and the key size is also 128 bits (16 characters). A comparison of the proposed encryption method with DES and AES is shown in table. 2. The asymmetric RSA algorithm is developed by MIT professors: Ronald L. Rivest, Adi Shamir, and Leonard M. Adleman in 1977 [5]. RSA gets its security from factorization problem. Difficulty of factoring large numbers is the basis of security of RSA.

In this Paper the actual message to be sent is encrypted and decrypted using the FSET algorithm which has been modified accordingly for higher efficiency. RSA is used for encryption and decryption of the secret key which is used in the encryption (FSET) of the actual data to be transmitted. All the limitations in FSET are overcome in this implementation. The security of the secret key is handled by the by the RSA. Here the FSET can handle multimedia data also. Multimedia files like images, videos, audios etc. can be effectively encrypted. Also other files like MS word, PDF, almost all files can be transmitted securely using the FSET proposed. The hash function is used for providing authentication. The detailed implementation is explained in the later sections.

## 2. THE ENCRYPTION ALGORITHM

An encryption algorithm has the advantages of both the symmetric and asymmetric algorithms. The complete process can be viewed in the figure 1. This process involves the fallowing steps

Step 1:- Finding Hash of message (hm) using SHA-2 (H)
Step 2:- Encryption of message using FSET with Secret key (Ks)
Step 3:- Perform XOR operation on hm and ks
Step 4:- Using RSA Encrypting the secret key and hnk (output of step 3) with Public key PU= {e, n}.
Step 5:- Using RSA, Decryption of hmk and encrypted secret key using Private Key PR= {d, n}
Step 6:- Perform XOR operation on Ks and hmk to get hm
Step 7:- Decryption of Message using FSET with Ks
Step 8:- Finding the Hash of decrypted message and comparing it with hm (output of step 6) to authenticate the message

## 3. THE ENCRYPTION PROCESS

The encryption process starts with the key generation process at the receiver side. The receiver generates two keys public and private key. The public key is sent to the sender and it is not necessarily to be kept secret. The secret key is used by the sender to encrypt the original message. The sender then generates hash code of original message and XOR with Secret key to get 'hmk' value. The sender then uses the public key and encrypts the secret key and 'hmk' using RSA. He then sends the encrypted message, encrypted hmk and the encrypted secret key to the receiver. The receiver first decrypts the secret key and hmk using RSA with private key. The secret key must be decrypted first as the encrypted message can only be decrypted with the original secret key. After the secret key is decrypted it is then used in the FSET algorithm to get back the original message using the FSET decryption algorithm. All the procedure is explained clearly in the fallowing sub sections.

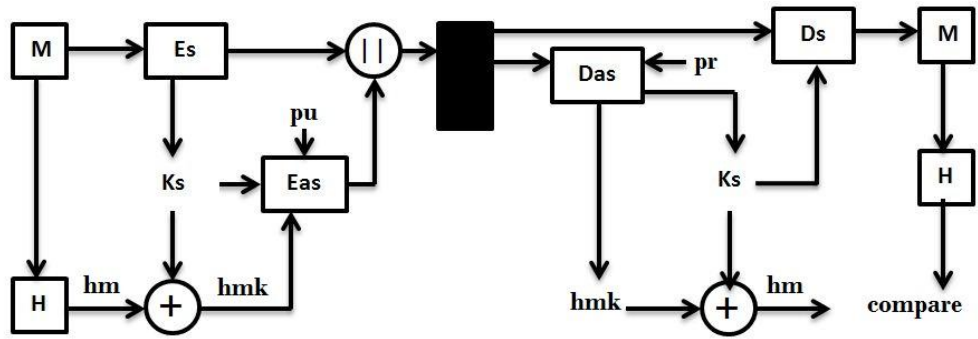

Figure 1: Implementation of the Algorithm

## 3.1 THE KEY GENERATION PROCESS

RSA involves a public key and a private key. The public key can be known to everyone and is used for encrypting messages. Messages encrypted with the public key can only be decrypted using the private key [3]. The keys for the RSA algorithm are generated the following way:

1. Choose two distinct large random prime numbers and
2. Compute n=p*q  n is used as the modulus for both the public and private keys
3. Compute the totient: f (n)=(p-1)(q-1).
4. Choose an integer e such that $1 < e < f(n)$ , and  share no factors other than 1 (i.e. e and φ(n) are co-prime)   e is released as the public key exponent
5. Compute d to satisfy the congruence relation de=1(mod f (n)); i.e. de=1+kf (n) for     some integer. d is kept as the private key exponent

The public key consists of the modulus and the public (or encryption)     exponent. The private key consists of the modulus and the private (or decryption) exponent which must be kept secret. Recipient after calculating public key PU= {e , n} and private key PR= {d, n} sends the public key value i.e., PU= {e ,n} value to sender.

## 3.2 SECRET KEY AND HMK ENCRYPTION USING RSA ALGORITHM

Receiver B transmits his public key to Sender and keeps the private key secret. Sender then wish to encrypt message M and hmk . He first turns M and hmk into a number m<n and hmk<n by using an agreed-upon reversible protocol known as a padding scheme. He then computes the cipher text corresponding to:

$$c=m^e \bmod n$$

This can be done quickly using the method of exponentiation by squaring. Sender then transmits to Receiver. Hmk is obtained by XOR operation on hash code obtained by performing hash function on message and the secret key.

## 3.3 FSET Encryption Algorithm and Hashing of Message

The encryption, $C = E(K,P)$, using the proposed encryption algorithm consists of three steps.

1. The first step involves initialization of a matrix with ASCII code of characters, shuffled using a secret key, $K$. This initialization is required only once before the beginning of conversion of a plaintext message into corresponding cipher text message.
2. The second step involves mapping, by substitution using the matrix, each character in every block of 16 characters into level-one cipher text character.
3. The third step involves translation and transposition of level-one cipher text characters within a block, by X-OR and circular shift operations, using arrays, in 8 rounds.

Figure 2 shows simplified block diagram of the encryption and decryption scheme.

### 3.3.1 Matrix for Substitution Mapping

A matrix $M$ with 16 rows and 256 columns initialized with ASCII codes of characters using secret key is used for mapping the plaintext characters into level one cipher text characters. During encryption, a block of 16 plaintext characters in the message is taken into a buffer. The ASCII code of the character $P(i)$ is obtained. The resulting integer is used as column number j of $i^{th}$ row of the matrix $M$. The element contained in this cell which is an ASCII code of a character, is taken as the level-one cipher text character $CL1(i)$ corresponding to the plaintext character $P(i)$. In this way all the characters in a block are mapped into level-one cipher text characters and all plaintext character blocks are

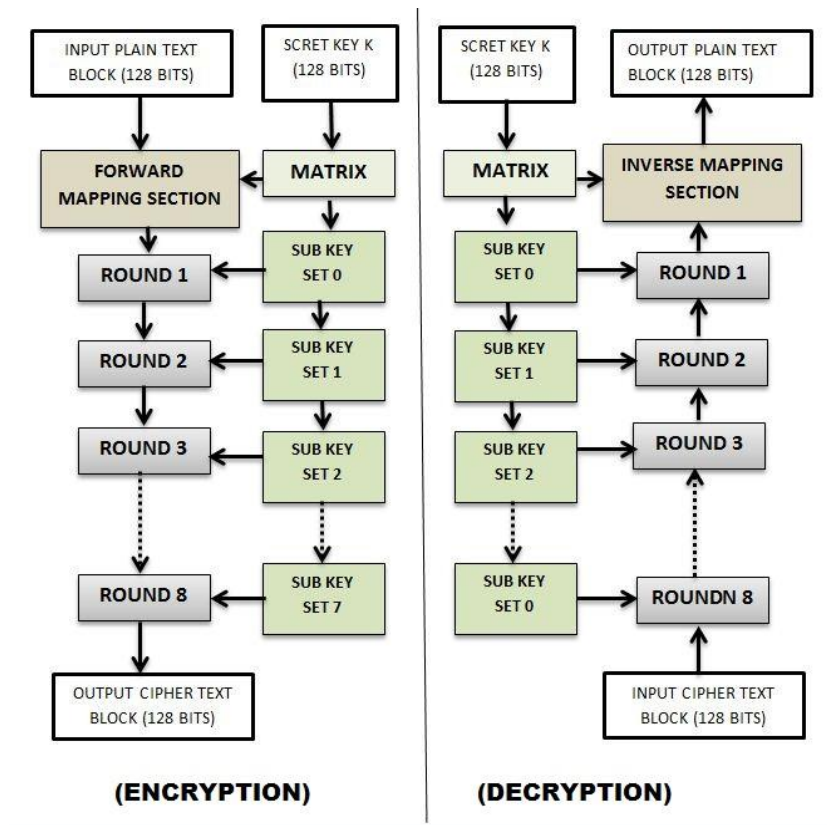

Figure 2: Block Diagram of Encryption & Decryption

### 3.3.2 MATRIX INITIALIZATION

A matrix *M* with sixteen rows and two hundred fifty six columns is defined. Columns in every row of the matrix is filled with ASCII codes of characters starting from NULL (ASCII = 0) in column zero to BLANK (ASCII = 255) in column two hundred fifty five representing elements of the matrix. A 16 character (128 bits) secret key *K*, with key characters *K(0)* through *K(15)*, is used for encryption and decryption. The ith row of the matrix is given an initial right circular shift, as many number of times as equal to the ASCII code of (i+1)th key character to shuffle the contents of the matrix *M,* for i = 0 to 14. For example, if *K(1)*, is .a. whose ASCII code is 97, row 0 of the matrix *M* is right circular shifted 97 times. If *K(2)* is .h. whose ASCII code is 104, the second row of the matrix *M* is right circular shifted 104 times and so on. The row 15 of matrix *M* is right circular shifted as many number of times as equal to ASCII value of the key character *K(0)*.

Further, the ith row of the matrix is given a second right circular shift as many number of times as equal to ASCII (*K(i)*) to shuffle the contents of the matrix *M*, for i = 0 to 15. For example, the row 0 of M is right circular shifted as many number of times as equal to the ASCII value of key character *K(0)*. The row 1 of the matrix *M* is given a right circular shift as many number of times as equal to the ASCII value of the key character *K(1)* and so on.

### 3.3.2 SUBSTITUTION MAPPING PROCEDURE

A given message is broken into blocks of sixteen plaintext characters *P(0)* through *P(15)*. Plaintext character *P(i)* is taken and a number j is calculated such that j = ( ASCII code of plaintext character *P(i))*. This number, j, is used as column number of the matrix *M*. Using j as column number we proceed to find the element in the $i^{th}$ row of the matrix *M*. This element (ASCII code of a character) is used as level-one cipher text character *CL1(i)* for a given plaintext character *P(i)*. For example, for the plaintext character *P(0)* in a block, i = 0, j = ( ASCII code of plaintext character *P(0))* is used as column number of row 0 of the matrix *M* to obtain level-one cipher text character corresponding to *P(0)*. Similarly for character *P(1)* in the plaintext character block, i = 1 and j = ( ASCII code of plaintext character *P(1))* where j is used as column number of the row 1 of the matrix to obtain level-one cipher text character corresponding to *P(1)*. In this way, all the 16 plaintext characters in a block are mapped into 16 level one cipher text characters denoted by *CL1(i)*, i = 0 to 15. The characters of level 1 cipher text character block (*CL1(0)* through *CL1(15)*) are transferred to a 16 element array *A1*.

### 3.3.3 SUB-KEY SET GENERATION

One set of eight sub-keys *Kts_0, Kts_1, Kts_2, .. Kts_7* are generated using the secret key *K* such that: *Kts_n* = characters in columns 0 through column 15 in row *n* of matrix *M concatenated*. These keys are used in translation rounds. Another set of sub-keys *Ktp_n0, Kps_n1, Ktp_n2* and *Ktp_n3 are* generated such that *Ktp_n0* = character of matrix *M* with row number n and column number 0. Here, each key is a character represented by the corresponding element in the matrix *M*. These keys are used in transposition rounds.

### 3.3.3 TRANSLATION AND TRANSPOSING

Eight rounds of translation and transposition operations are performed on the level 1 cipher text character block. The translation operations are done using XOR operation performed on the cipher text character block using sub key, *Kts_n* in the nth round. The translated cipher text character block is transposed using four arrays whose elements are circular shifted using sub-keys *Ktp_n0, Ktp_n1, Ktp_n2, Ktp_n3* used in that round. These operations make the resulting output cipher text characters extremely difficult to decrypt by any adversary without having the secret key. The translation and transposition produce the effect of diffusion.

**Translation of cipher text characters**

The contents of array *A1* is XOR with sub key *Kts_n* in the nth round. The 16 characters of each block of cipher text are XOR with 16 characters of sub key *Ks_n*

**Transposing of cipher text characters**

The XOR level-one cipher text characters available in array *A1* are bifurcated and transposed using four arrays. For the nth round, array *A1* is right circular shifted as many number of times as equal to the integer value of *Ktp_n0*. After this operation, the first eight elements of *A1* (left most elements) are transferred to another array *A2* having 8 element positions. Then, *A2* is right circular shifted as many number of times as equal to the integer value of *Ktp_n1*. The other eight elements of the array *A1* (rightmost elements) are transferred to another 8 element array *A3* which is left circular shifted as many number of times as equal to integer value of *Ktp_n2*. Then *A2* and *A3* are concatenated and transferred to the 16 element array *A1*. This 16 element array, *A1*, is right circular shifted as many number of times as equal to the integer value of *Ktp_n3*. After this operation, the contents of *A1* represent the cipher text characters in a given block. The elements of array *A1* are moved to the cipher text block *C(0)* through *C(15)*. The cipher text blocks are used to create the output cipher text message file.

### 3.3.4 HASHING OF MESSAGE

At this stage hash code of the message is found using the Secure Hash Algorithm (SHA -2)

## 3.4 SECRET KEY AND HMK DECRYPTION USING RSA ALGORITHM

Receiver b can recover m and hmk by using her private key exponents d by the following computation:

$$M = C^d \bmod n.$$

Given m, he can recover the original message key Ks and hmk. Here M may original be hmk and Ks and C is their respective encrypted form. Once hmk is decrypted it is then XOR-ed with Ks to get hm.

## 3.5 THE DECRYPTION PROCESS AND AUTHENTICATION

The decryption algorithm performs the reverse operations of encryption such that $P = D(K,C)$. This is done in three steps. Here, cipher text character $C(i)$, in blocks of 16 are processed using arrays and matrix. The first step involves initialization of a matrix with ASCII codes of characters, shuffled using the secret key. In the second step, the cipher text characters are de-transposed using circular shift operation of array and de-translated by XOR logic using sub-keys in multiple rounds. With this operation we get back the level-one cipher text characters. In the third step, these level-one cipher text characters are inverse-mapped into plaintext characters using the matrix. In the decryption algorithm, sub-keys are generated from the secret key in the same way as in the case of encryption algorithm. The detailed procedure is explained in the fallowing sections.

### 3.5.1 MATRIX INITIALIZATION

An identical matrix *M*, used for mapping the plaintext characters into level-one cipher text characters, is used here for inverse mapping of the level-one cipher text characters into plaintext characters during decryption. At the decryption site, this matrix is created using the secret key *K* in the same way as in the case of encryption.

### 3.5.2 DE-TRANSPOSING OF CIPHER TEXT CHARACTERS

The cipher text character block from the cipher text file is brought in to a 16 element array *A1*. For the nth round, array *A1* is left circular shifted as many number of times as equal to the integer value of $Ktp\_n3$. After this operation, the first eight elements of *A1* (left most elements) are transferred to another array *A2* having 8 element positions. Then, *A2* is left circular shifted as many number of times as equal to the integer value of $Ktp\_n2$. The other eight elements of the array *A1* (rightmost elements) are transferred to another 8 element array *A3* which is right circular shifted as many number of times as equal to integer value of $Ktp\_n1$. Then *A2* and *A3* are concatenated and transferred to the 16 element array *A1*. This array is left circular shifted as many number of times as equal to the integer value of $Ktp\_n0$.

### 3.5.3 DE-TRANSLATION OF CIPHER TEXT CHARACTERS

The contents of array *A1* is X-ORed with the bits of sub key $Kts\_n$ in the nth round. After this operation, the contents of the array *A* corresponds to the level one cipher text character block corresponding to the one obtained after the mapping operation done at the encryption side using the matrix. The contents of array *A1* is moved to level 1 cipher text block, *CL1*.

### 3.5.4 INVERSE MAPPING USING MATRIX

If *CL1(i)* is the level-one cipher text character in a block, the inverse mapping is such that *P(i)* = char((column number j of $i^{th}$ row of matrix *M* where *CL1(i)* is the element)). For example, let the 1st level-one cipher text character, *CL1(1)*, in a block be .#.. We proceed to search .#. in the matrix *M* to find the column number j in the 1st row where *CL1(1) = M[1][j]*. Then we determine the character whose ASCII = (j) which gives the plaintext character *P(1)* corresponding to *CL1(1)*. Let the $2^{nd}$ level-one cipher text character, *CL1(2)*, in a block be .%.. We proceed to search .%. in the matrix *M* to find the column number j in the 2nd row where *CL1(2) = M[2][j]*. Then we determine the character whose ASCII = (j) which gives the plaintext character *P(2)* corresponding to *CL1(2)*. In this way we can inverse map every cipher text character in every block into plaintext characters to get back the original message file.

### 3.5.5 AUTHENTICATION PROCEDURE

Once the message is decrypted, hash function (SHA-2) is used to calculate the hash code of the original message and it is compared with hm. The message is authentic if the comparison result is true. If the result of comparision of result is false the authentication fails.

## 4. SIMULATION AND EXPERIMENTAL RESULTS

In this section we have shown the encryption of an image file. The key generation process can be seen in the figure 3. It shows the selected prime numbers and generated public and private key values. A secret key is chosen "encryption algorithm" which can be seen in the figure 4.

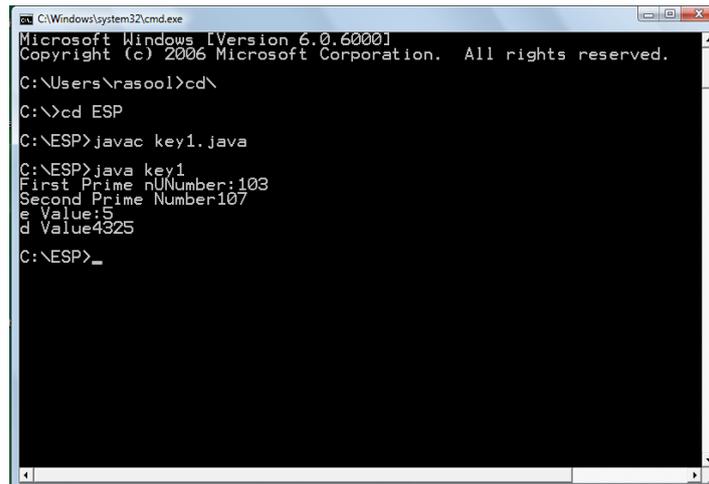

Figure 3: Key Generation process

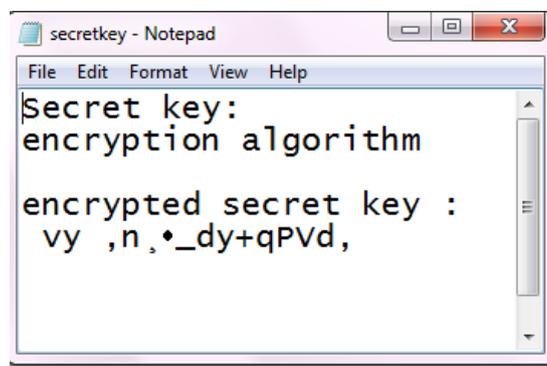

Figure 4: Secret key

The public key is used in RSA algorithm to encrypt the secret key file and hmk. The encrypted secret key can be seen in figure 4. The secret key is used for encrypting the image file suing FSET algorithm. The encrypted image cannot be opened. It's highly secure.

Performance comparison of various popular secret key algorithms, such as DES, AES and Blowfish running on a Pentium-4, 2.4 GHz machine, discussed in the literature [9] shows that Blowfish is the fastest among these algorithms. The throughputs of these algorithms are respectively 4,980 bytes/sec, 2,306 bytes/sec and 5,167 bytes/sec. The proposed FSET Symmetric-key Encryption algorithm is subjected to performance evaluation using a Pentium-4, 2.4 GHz machine. Execution time taken by the algorithm was measured using a image file and the throughput calculated. The time between two test points in the algorithm during execution was measured with the help of system clock.

The number of bytes (in the plaintext file) required for an execution time of one second during encryption was ascertained. The comparison of performance of this encryption algorithm with the performance of popular secret key algorithms given in [4] is made. The throughput of Blowfish algorithm is only 5,167 bytes per second whereas FSET encryption algorithm provides 70,684 bytes per second. Thus this Encryption algorithm is 8 times faster than Blowfish algorithm.

## 5. CONCLUSION

The proposed encryption technique has the advantages of both symmetric and asymmetric algorithms. Symmetric algorithm is used for encryption of messages rather than asymmetric because the asymmetric algorithms are slower compared to symmetric algorithms. Thus Asymmetric algorithm RSA is used here to safeguard the secret key which solves the problem of key exchange as the secret key can be sent securely. The secret key can't be decrypted unless a private key is obtained and since it is at receiver side it is highly secured. Hashing provides authentication to the messages sent.

The FSET Encryption algorithm, presented above, is a simple, direct mapping algorithm using matrix and arrays. Consequently, it is very fast and suitable for high speed encryption applications. The matrix based substitution resulting in poly alphabetic cipher text generation followed by multiple round arrays based transposing and XOR logic based translations give strength to this encryption algorithm. The combination of poly alphabetic substitution, translation and transposition makes the decryption extremely difficult without having the secret key. Decryption of cipher text messages created using this encryption is practically impossible by exhaustive key search as in the case of other algorithms using 128 bits secret key. The cipher text generated by this algorithm does not have one to one correspondence in terms of position of the characters in plaintext and cipher text. This feature also makes decryption extremely difficult by brute force. The performance test shows that this encryption is a fast algorithm compared to the popular Symmetric-key algorithms. The algorithm is enhanced so that it can handle various kinds of data like images, videos, PDF etc.

**Authors**

**Shaik Rasool**[1] received the Bachelor of Technology in Computer Science & Engineering from Jawaharlal Nehru Technological University, Hyderabad, India in 2008. He is currently pursuing Master of Technology in Computer Science & Engineering from Jawaharlal Nehru Technological University and also working as Assistant Professor at the Department of Computer Science & Engineering in S.C.E.T., Hyderabad, India. His main research interest includes Network Security, Biometrics, Data Mining, and Information Security, Programming Language and security and Artificial Intelligence.

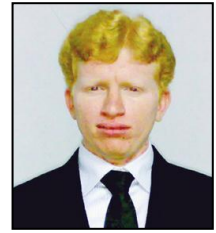

**G. Sridhar**[2] received his B.S. in Computer Science & Information Technology and M.S. in Computer Science and Information Technology from State Engineering University of Armenia, Yerevan, Armenia. He is currently working as Associate Professor at the Department of Computer Science & Engineering in S.C.E.T., Hyderabad, India. His main research interest includes Information Security, Software Testing Methodologies and Software Models.

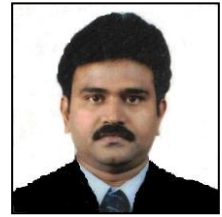

**K. Hemanth Kumar**[3] received the Bachelor of Technology in Computer Science & Engineering from Jawaharlal Nehru Technological University, Hyderabad, India in 2005 and Master of Technology in Computer Science & Engineering from Jawaharlal Nehru Technological University, Kakinada, India in 2010 and also working as Assistant Professor at the Department of Computer Science & Engineering in S.C.E.T., Hyderabad, India. His main research areas are Information Security and Computer Networks.

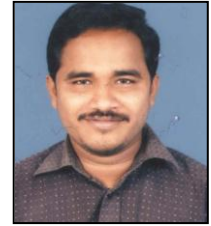

**P. Ravi Kumar**[4] received the Master of Technology in Computer Science & Engineering from Bharat University, Chennai, India in 2008 and also working as Assistant Professor at the Department of Computer Science & Engineering in S.C.E.T., Hyderabad, India. His main research interest includes Data Mining, and Information Security.

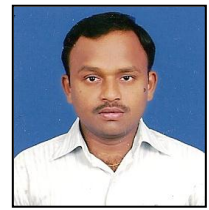